\documentclass[a4paper]{jpconf}
\usepackage{graphicx}
\usepackage[mathscr]{eucal}
 \usepackage{amsmath, amssymb}
\def\dmf{\dot{\mathfrak{M}}}
 \def\msE{\mathscr{E}}
   \def\sun{\hbox{$\odot$}}

\newcommand{\be}{\begin{equation}}
\newcommand{\ee}{\end{equation}}
\newcommand{\bdm}{\begin{displaymath}}
\newcommand{\edm}{\end{displaymath}}
\begin{document}

\vspace{-5.5cm}
\noindent {\it Journal of Physics: Conference Series, 2012}\\
Proc. of AHAR\,2011 Conference ``The Central Kiloparsec in Galactic Nuclei'', Bad-Honnef, Germany, Aug. 29 -- Sep. 2, 2011

\vspace{2cm}

\title{Magnetically controlled accretion onto a black hole}

\author{N R Ikhsanov$^1$, L A Pustil'nik$^2$ and N G Beskrovnaya$^1$}

\address{$^1$~Pulkovo Observatory, Pulkovskoe Shosse 65, Saint-Petersburg 196140, Russia}
\address{$^2$~Israel Space Weather and Cosmic Ray Center, Tel Aviv University, Israel Space Agency, \& Golan Research Institute, Israel}

\ead{ikhsanov@gao.spb.ru}

\begin{abstract}
An accretion scenario in which the material captured by a black hole from its environment is assumed to be magnetized ($\beta \sim 1$) is discussed. We show that the accretion picture in this case is strongly affected by the magnetic field of the flow itself. The accretion power within this Magnetically Controlled Accretion (MCA) scenario is converted predominantly into the magnetic energy of the accretion flow. The rapidly amplified field prevents the accretion flow from forming a homogeneous Keplerian disk. Instead, the flow is decelerated by its own magnetic field at a large distance (Shvartsman radius) from the black hole and switches into a non-Keplerian dense magnetized slab. The material in the slab is confined by the magnetic field and moves towards the black hole on the time scale of the magnetic field annihilation. The basic parameters of the slab are evaluated. Interchange instabilities in the slab may lead to a formation of Z-pinch type configuration of the magnetic field over the slab in which the accretion power can be converted into jets and high-energy radiation.
\end{abstract}

\section{Introduction}
We consider a black hole of the mass $M_{\rm bh}$, surrounded by a gas of the average density $\rho_{\infty}$, which moves in the frame of the black hole with a relative velocity $v_{\rm rel} = \left(v_{\rm s}^2 + v_{\rm w}^2\right)^{1/2}$, where $v_{\rm s}$ and $v_{\rm w}$ are the sound speed and the velocity of the gas proper motion. A maximum distance at which the black hole interacts with its environment through gravitational potential is the Bondi radius $R_{\rm G} = 2 GM_{\rm bh}/v_{\rm rel}^2$. It is defined by equation $v_{\rm p}(R_{\rm G}) = v_{\rm rel}$, where $v_{\rm p}(r) = \left(2GM_{\rm bh}/r\right)^{1/2}$ is the parabolic (free-fall) velocity at a distance $r$. The mass with which the black hole interacts in a unit time is $\dmf_{\rm c} = \pi R_{\rm G}^2 \rho_{\infty} v_{\rm rel}$. The mass capture rate by the black hole from its environment in general case is limited to $\dmf_{\rm a} \leq \dmf_{\rm c}$.

The structure of the accretion flow onto a black hole and the mode by which the accretion power is released depends on the initial conditions in the material captured by the black hole at the Bondi radius. A situation in which this material is slowly moving, i.e. $v_{\rm w} < v_{\rm s}$, and non-magnetized is discussed in the next section. It can be expected if the black hole captures material from a giant molecular cloud. If the material captured by the black hole is supplied by nearby stars its properties can significantly differ from those mentioned above. The velocity of stellar wind usually exceeds the sound speed in the outflowing material, i.e. $v_{\rm w} \gg v_{\rm s}$, and the magnetic pressure in the wind, $\msE_{\rm m}^{(0)} = B_{\rm w}/8 \pi$, is comparable to its thermal pressure, $\msE_{\rm th}^{(0)} = \rho_{\infty} v_{\rm s}^2$. This is justified by studies of solar wind (see e.g. \cite{Stumik-etal-2011}, and references therein) and recent results on the surface field of massive O/B-type stars (see e.g. \cite{Walder-etal-2011}, and references therein). The accretion picture which is realized under these conditions is discussed in Sect.\,\ref{mca}. We show that accretion picture in this case is strongly affected by the magnetic field of the flow itself. It prevents the material from forming a Keplerian disk. Instead appearance of a dense magnetized non-Keplerian slab is expected. Formation of the slab can under certain conditions result in a strong electric potential in the vicinity of the black hole which might be responsible for powerful jets. Our conclusions are summarized in Sect.\,\ref{conclusions}.

  \section{Non-magnetized accretion flow scenario}

Accretion scenario in AGNs is usually built around the assumptions that the material surrounding the black hole is non-magnetized and moves with a relatively small velocity, i.e. $v_{\rm rel} \sim v_{\rm s}$. As this material is captured by a black hole it initially follows ballistic trajectories forming a quasi-spherical accretion flow. If the accreting material possesses angular momentum the flow can switch its geometry from quasi-spherical to a Keplerian disk. For the disk to form the circularization radius, $R_{\rm circ} = \dot{J}^2/GM_{\rm bh} \dmf_{\rm a}^2$, at which the angular velocity of the material, $\omega_{\rm en} = \xi \Omega_0 \left(R_{\rm G}/r\right)^2$, reaches the Keplerian angular velocity, $\omega_{\rm k} = \left(2GM_{\rm bh}/r^3\right)^{1/2}$, should exceed a distance at which the ballistic trajectories of the material are truncated. Here $\dot{J} = \xi j_0 \dmf_{\rm a}$ is the angular momentum accretion rate, $j_0$ is the specific angular momentum of the material captured by a black hole at the Bondi radius, and $\xi$ is the factor by which angular momentum accretion rate is reduced due to inhomogeneities (velocity and density gradients) and the magnetic viscosity in the accretion flow. The specific angular momentum carried by a material captured by the black hole from turbulent Interstellar Medium (ISM) can be evaluated as $j_0 = v_{\rm t}(R_{\rm t}) R_{\rm t}^{-1/3} R_{\rm G}^{4/3}$, where $v_{\rm t}(R_{\rm t})$ is the velocity of turbulent motions at the scale of $R_{\rm t}$ and the Kolmogorov spectrum of the turbulent motions is assumed (see e.g. \cite{Prokhorov-etal-2002}, and references therein). The circularization radius of the accretion flow in this case can be expressed as
 \be\label{rcirc}
R_{\rm circ} = \frac{\xi^2 v_{\rm t}^2 R_{\rm t}^{-2/3} R_{\rm G}^{8/3}}{GM_{\rm bh}}.
 \ee
A formation of Keplerian accretion disk would be expected if $R_{\rm circ} > \kappa_{\rm r} R_{\rm g}$, where $R_{\rm g} = 2GM_{\rm bh}/c^2$ is the gravitational radius of the black hole, $\kappa_{\rm r} = R_{\rm st}/R_{\rm g}$, and $R_{\rm st}$ is the radius of the last stable orbit around the black hole. Solving this inequality for $v_{\rm rel}$ yields $v_{\rm rel} < v_{\rm cr}^{(0)}$, where
 \be
v_{\rm cr}^{(0)} \simeq 690\,{\rm km/s}\ \times\ \kappa_{\rm r}^{-3/16} m_{6}^{1/8} \left(\frac{v_{\rm t}}{10\,{\rm km/s}}\right)^{3/8} \left(\frac{R_{\rm t}}{10^{20}\,{\rm cm}}\right)^{-1/8}.
 \ee
Here $m_6$ is the mass of the black hole in units of $10^6\,{\rm M_{\sun}}$, and parameters $v_{\rm t}$ and $R_{\rm t}$ are normalized following \cite{Ruzmaikin-etal-1998}. The derived value of $v_{\rm cr}^{(0)}$ significantly exceeds the sound speed in ISM and thus, a supermassive black hole within the non-magnetized accretion flow scenario is expected to be surrounded by a Keplerian accretion disk.

As a black hole is accreting material from the Keplerian disk its gravitation energy is converted predominantly into the kinetic and thermal energy of the accreting material, which then can be converted into the magnetic energy due to dynamo action. The pressure of the magnetic field generated in the disk? therefore, is limited to $\msE_{\rm m}^{\rm (d)}(r) < \rho(r) v_{\rm k}^2(r)$, where $v_{\rm k}(r) = \left(GM_{\rm bh}/r\right)^{1/2}$ is the Keplerian velocity. An influence of the field on the accretion flow in this case is relatively small and does not lead to any significant changes in the flow dynamics. The disk remains Keplerian and the accreting material approaches the black hole on the viscous timescale. The scale of the magnetic field generated in the turbulent differentially rotating Keplerian disk by the dynamo action is limited to the disk thickness (which represents the largest scale of the turbulent motions in the disk). The energy of the magnetic field is released in the disk corona, which is formed as the field emerges from the disk due to the buoyancy instability. The energy release process is associated with the magnetic reconnection which leads to heating and ejection of material \cite{Galeev-etal-1979}. However, the kinetic luminosity of the ejecta is relatively small and the outflowing material is non-collimated.

Existence of collimated powerful jets cannot be explained within this scenario unless some additional assumptions are incorporated into the model. In particular, one can assume that the jets are powered by the rotational energy of the black hole, which is released as the inner radius of the disk is approaching the last stable orbit. The ejected material can be self-collimated, or (which looks more reasonable) is collimated by the large-scale magnetic field of the disk. The field in the latter case should be strong enough to collimate a powerful jet. This criterium is difficult to satisfy if the large-scale field of the disk is provided by the small-scale magnetic arches generated in the disk due to dynamo action. The pressure of the magnetic field in this situation is substantially smaller than the ram pressure of the accretion flow itself. It is more likely that the large-scale field of the disk was initially present in the material captured by the black hole and has been amplified during the accretion process. In the next Section we show that this assumption cannot be simply incorporated into the traditional accretion model. The magnetic field in the spherical accretion flow increases rapidly and prevents accreting material from forming a Keplerian disk.

 \section{Magnetically controlled accretion scenario}\label{mca}

A situation in which $v_{\rm rel} \approx v_{\rm w}$ and $\beta = \msE_{\rm th}^{(0)}/\msE_{\rm m}^{(0)} \sim 1$ can be realized if the material captured by a black hole is supplied by the stellar wind of nearby stars. The ram pressure of the material at the Bondi radius, $\msE_{\rm r}^{(0)} = \rho_{\infty} v_{\rm w}^2$, in this case significantly exceeds its thermal and magnetic pressure and hence, the Alfv\'en velocity, $v_{\rm A} = B_{\rm w}/(4 \pi \rho)^{1/2}$, in the accreting material is much smaller than the free-fall velocity. The time of the magnetic field annihilation in the accretion flow,
  \be\label{trec}
 t_{\rm rec} = \frac{r}{\eta_{\rm m} v_{\rm A}} = \eta_{\rm m}^{-1} t_{\rm ff}
 \left(\frac{v_{\rm ff}}{v_{\rm A}}\right),
 \ee
under these conditions significantly exceeds the dynamical (free-fall) time, $t_{\rm ff} = \left(r^3/2GM_{\rm bh}\right)^{1/2}$, and the magnetic flux in the free-falling gas is conserved. Here $B_{\rm w}$ is the field strength in the accretion flow and the efficiency parameter of the magnetic reconnection ranges in the interval 0.01--0.15 (see e.g. \cite{Parker-1971, Noglik-etal-2005}).

 \subsection{Shvartsman radius}

The magnetic field in the free-falling material is dominated by the radial component \cite{Zeldovich-Shakura-1969}, which under the condition of the magnetic flux conservation increases as $B_{\rm r}(R) \sim B_{\rm f}(R_{\rm G}) \left(r/R_{\rm G}\right)^{-2}$ \cite{Bisnovatyi-Kogan-Fridman-1970}. The magnetic pressure in the accreting material, therefore, increases while approaching the black hole as
\be
 \msE_{\rm m}(r) = \msE_{\rm m}(R_{\rm G}) \left(\frac{r}{R_{\rm G}}\right)^{-4},
 \ee
and the ram pressure of the free-falling spherical flow is
 \be\label{eram}
\msE_{\rm ram}(r) = \msE_{\rm ram}^{(0)} \left(\frac{r}{R_{\rm G}}\right)^{-5/2}.
 \ee
This indicates that the magnetic energy in the free-falling gas increases more rapidly than its kinetic energy,  $\msE_{\rm m}/\msE_{\rm ram} \propto r^{-3/2}$, and hence, the gravitational energy of the black hole in the scenario under consideration is converted predominantly into the magnetic energy of the accreting material.

A distance $R_{\rm sh}$, at which the magnetic pressure in the accretion flow reaches its ram pressure, can be evaluated by equating $\msE_{\rm m}(R_{\rm sh}) = \msE_{\rm ram}(R_{\rm sh})$. This yields \cite{Shvartsman-1971}
 \be\label{rsh}
 R_{\rm sh} = \beta^{-2/3} \left(\frac{V_{\rm s}}{V_{\rm rel}}\right)^{4/3} R_{\rm G} =
 \beta^{-2/3}\ \frac{2 GM_{\rm bh} V_{\rm s}^{4/3}}{V_{\rm rel}^{10/3}}.
 \ee
Parameter $R_{\rm sh}$ (hereafter Shvartsman radius) represents a minimum distance from which the accretion process is fully controlled by the magnetic field of the flow itself. The Alfv\'en velocity in the accretion flow at this distance reaches the free-fall velocity. An accretion of homogeneous gas in which the magnetic flux is conserved inside Shvartsman radius is impossible. Otherwise, the magnetic energy in the flow would exceed the gravitational energy, which contradicts the energy conservation law (for discussion see \cite{Shvartsman-1971}). Further accretion, therefore, can be realized only on the timescale of the field dissipation, $t_{\rm rec}$, which is significantly larger than the free-fall time (see Eq.~\ref{trec}). This indicates that the initially free-falling accretion flow is decelerated by its own magnetic field at $R_{\rm sh}$. The deceleration leads to formation of a shock in which the gas is heated up to a temperature $T_{\rm s} = (3/16) T_{\rm ff}(R_{\rm sh})$, where $T_{\rm ff}(r) = GM_{\rm bh} m_{\rm p}/k_{\rm B} r$ is the proton free-fall temperature, $m_{\rm p}$ is the proton mass and $k_{\rm B}$ is the Boltzmann constant \cite{Lamb-etal-1977}. If cooling of the material in the region of flow deceleration is ineffective a hot envelope surrounding the black hole at the Shvartsman radius forms.

Rapid amplification of the magnetic field in the spherical flow as well as deceleration of the flow by its own magnetic field at the Shvartsman radius have been confirmed by the results of numerical studies of magnetized spherical accretion onto a black hole \cite{Igumenshchev-etal-2003, Igumenshchev-2006}. These calculations have shown that the magnetized flow under the conditions of interest is shock-heated at the Shvartsman radius up to the adiabatic temperature. These authors have considered further accretion under assumptions that the flow is radiatively inefficient and heating of the flow dominates cooling. An overheating of material in this case can lead to transition of the accretion flow into convective-dominated stage in which a portion of the accreting material is leaving the system in a form of turbulent jets and the mass accretion rate inside the region of the flow deceleration significantly decreases. Here we show, however, that MCA picture can be constructed without these assumptions and a question about the nature of energy source responsible for the overheating of material (which is required for the flow to switch into the convective-dominated phase) can be avoided.

  \subsection{No Keplerian disk within MCA scenario}

Ballistic trajectories of the free-falling spherical accretion flow within MCA scenario are truncated at the Shvartsman radius. The condition for the Keplerian disk formation in this case reads $R_{\rm circ} \geq R_{\rm sh}$. This inequality can be expressed using  Eqs.~(\ref{rcirc}) and (\ref{rsh}) as $v_{\rm rel} \leq v_{\rm cr}^{\rm mag}$, where
  \be
 v_{\rm cr}^{\rm mag} \simeq 5 \times 10^5\,{\rm cm\,s^{-1}}\ \beta^{1/3} \xi_{0.2} m_6^{1/3} \left(\frac{v_{\rm s}}{10\,{\rm km\,s^{-1}}}\right)^{-2/3} \left(\frac{V_{\rm t}}{10\,{\rm km\,s^{-1}}}\right) \left(\frac{R_{\rm t}}{10^{20}\,{\rm cm}}\right)^{-1/3}.
  \ee
Parameter $\xi_{0.2} = \xi/0.2$ is normalized here to its maximum average value, which has been derived in numerical studies of the accretion process within the non-magnetized spherical accretion flow approximation \cite{Ruffert-1999}. Since the derived value of $v_{\rm cr}^{\rm mag}$ is comparable or even smaller than the sound speed in the surrounding gas the angular momentum of the accreting material at the Shvartsman radius appears to be insufficient for the Keplerian disk to form.

  \subsection{Cooling of the accretion flow}

Cooling of the accretion flow in the considered case is dominated by the inverse Compton scattering of electrons on photons emitted in the vicinity of the black hole. The Compton cooling time of the material at $R_{\rm sh}$ can be evaluated as \cite{Elsner-Lamb-1977},
 \be\label{tcomp}
t_{\rm c}(R_{\rm sh}) = \frac{3 \pi m_{\rm e} c^2 R_{\rm sh}^2}{2 \sigma_{\rm T} L_{\rm bh}},
 \ee
where $m_{\rm e}$ is the electron mass, $\sigma_{\rm T}$ is the Thomson cross-section and $L_{\rm bh}$ is the luminosity of the source associated with the black hole. Cooling is dominating heating at the Shvartsman radius if $t_{\rm c}(R_{\rm sh}) \leq t_{\rm ff}(R_{\rm sh})$. Solving this inequality yields $v_{\rm rel} \geq v_{\rm cc}$ where
 \be
 v_{\rm cc} \simeq 2 \times 10^7\ \beta^{-1/5} m_6^{3/5} L_{42}^{-3/5} \left(\frac{v_{\rm s}}{10\,{\rm km\,s^{-1}}}\right)^{2/3}\ {\rm cm\,s^{-1}},
 \ee
and $L_{42} = L_{\rm bh}/10^{42}\,{\rm erg\,s^{-1}}$. The value of $v_{\rm cc}$ is much smaller than typical wind velocity of massive stars and is even comparable to the velocity of stellar wind emitted by the red dwarfs. This indicates that a stationary magnetically controlled accretion at the rate $\dmf_{\rm a} \approx \dmf_{\rm c}$ is expected if the material captured by the black hole is predominantly supplied by the stellar wind of nearby stars. Compton cooling in this case prevents the flow from switching into the convective-dominated stage.

  \subsection{Non-Keplerian slab}

The accretion picture of a cold ($T < T_{\rm s}$) magnetized ($\beta < 1$) gas has been discussed in \cite{Bisnovatyi-Kogan-Ruzmaikin-1974, Bisnovatyi-Kogan-Ruzmaikin-1976}. It has been shown that the material in this case tends to flow along the magnetic field lines and in the region $r \leq R_{\rm sh}$ is accumulated in a dense non-Keplerian slab (see Fig.\,1 in \cite{Bisnovatyi-Kogan-Ruzmaikin-1976}). The material in the slab is confined by the magnetic field of the flow itself and its radial motion continues as the field is annihilating. The accretion process in the slab, therefore, occurs on the timescale of $t_{\rm rec}$.

The magnetic pressure inside the Shvartsman radius increases according to the energy conservation law as $\msE_{\rm m} = \msE_{\rm m}(R_{\rm sh}) \left(R_{\rm sh}/r\right)^{5/2}$, where
 \be
 \msE_{\rm m}(R_{\rm sh}) = \frac{\dmf_{\rm c} (2GM_{\rm bh})^{1/2}}{4 \pi R_{\rm sh}^{5/2}}
 \ee
is the magnetic pressure at the Shvartsman radius. This indicates that the field strength in the vicinity of the black hole reaches the value
 \be
 B(R_{\rm g}) = 20\ L_{42}^{1/2} m_6^{-1}\ {\rm kG}.
 \ee

Since the material is confined by the field in the slab its thermal pressure does not exceed the magnetic pressure. The gas number density at the inner radius of the slab in this case can be estimated as
  \be\label{rhosl}
  \rho_{\rm sl} \leq 10^{19}\,{\rm cm^{-3}}\ L_{42} m_6^{-2} T_4^{-1},
  \ee
where $T_4$ is the gas temperature at the inner radius of the slab in units of $10^4$\,K. The thickness of the slab depends on the magnetic field configuration and in the first approximation can be evaluated from continuity equation as
 \be\label{hz}
 h_{\rm z}(R_{\rm g}) \simeq 10^5\,{\rm cm}\ \eta_{0.01}^{-1} m_6 T_4,
 \ee
where $\eta_{0.01} = \eta_{\rm m}/0.01$. It should be noted, that the thickness of the slab strongly depends on
the field annihilation time, which can be evaluated from the shortest timescale of the source variability. Furthermore, we cannot discard a possibility that the slab is disintegrated by interchange instabilities into a large number of filaments connected by the field lines as it sketched in Fig.\,1 in \cite{Ikhsanov-Pustilnik-1994}. Formation of jets in this case can be associated with interaction between the magnetic arches connecting the filaments which leads to a Z-pinch field configuration over the slab. The process of energy release and plasma ejection in this configuration will be discussed in a forthcoming paper.

\section{Conclusions}\label{conclusions}

Black holes within the non-magnetized accretion flow scenario are expected to be surrounded by Keplerian disks. The scale of the magnetic field generated in the disk due to dynamo action is comparable to the disk thickness and the magnetic pressure associated with this field insignificantly contributes the total power of the disk. A formation of a strong large-scale magnetic field in the accretion flow can be expected if the material captured by the black hole is magnetized. The behavior of the accreting material in this case is strongly affected by its magnetic field. The flow is decelerated by the magnetic field at the Shvartsman radius and moves towards the black hole on the timescale of the magnetic field annihilation. Formation of a homogeneous Keplerian disk within this Magnetically Controlled Accretion (MCA) scenario does not occur. Instead, the accreting material is accumulated in a non-Keplerian dense slab. The material in the slab is confined by its own magnetic field and moves towards the black hole as the magnetic field is annihilating. Interchange instabilities of the slab may lead to formation of the Z-pinch type configuration of the magnetic field in which the accretion power is released in the form of jets and high energy emission.

\ack
We would like to thank G.S.\,Bisnovatyi-Kogan, M.V.\,Medvedev and P.L.\,Biermann for useful discussions. Nazar Ikhsanov is grateful to Alexander von Humboldt Foundation for support in attending the conference and Michael Kramer for useful discussions and kind hospitality at Max-Planck Institute of Radio Astronomy in Bonn. The research has been partly supported by the Program of RAS Presidium N\,19, NSH-3645.2010.2, and by the grant ``Infrastructure'' of Israel Ministry of Science.


\section*{References}
\begin{thebibliography}{9}
\bibitem{Bisnovatyi-Kogan-Fridman-1970}
 Bisnovatyi-Kogan, G.S., Fridman, A.M., 1970, \emph{Soviet Astronomy}, \textbf{13}, 566--568

\bibitem{Bisnovatyi-Kogan-Ruzmaikin-1974}
 Bisnovatyi-Kogan, G.S., Ruzmaikin, A.A. 1974, \emph{Astrophys. and Space Sci.}, \textbf{28}, 45--59

\bibitem{Bisnovatyi-Kogan-Ruzmaikin-1976}
  Bisnovatyi-Kogan, G.S., Ruzmaikin, A.A. 1976, \emph{Astrophys. and Space Sci.}, \textbf{42}, 401--424

\bibitem{Elsner-Lamb-1977}
 Elsner R.F., Lamb F.K. 1977, \emph{Astrophys. J.}, \textbf{215}, 897--913

\bibitem{Galeev-etal-1979}
 Galeev, A.A., Rosner, R., Vaiana, G.S. 1979, \textit{Astrophys. J.}, \textbf{229}, 318--326

\bibitem{Igumenshchev-etal-2003}
 Igumenshchev, I.V., Narayan, R., Abramowicz, M.A. 2003, \emph{Astrophys. J.}, \textbf{592}, 1042--1059

\bibitem{Igumenshchev-2006}
 Igumenshchev, I.V. 2006, \emph{Astrophys. J.}, \textbf{649}, 361--372

\bibitem{Ikhsanov-Pustilnik-1994}
 Ikhsanov, N.R., Pustil'nik, L.A. 1994, \emph{Astrophys. J. Suppl.}, \textbf{90}, 959--961

\bibitem{Lamb-etal-1977}
 Lamb, F.K., Fabian, A.C., Pringle, J.E., Lamb, D.Q. 1977, \emph{Astrophys. J.}, \textbf{217}, 197--212

\bibitem{Noglik-etal-2005}
 Noglik, J.B., Walsh, R.W., Ireland, J. 2005, \emph{Astron. Astrophys.}, \textbf{441}, 353--360

\bibitem{Parker-1971}
 Parker, E.N. 1971, \emph{Astrophys. J.}, \textbf{163}, 279--285

\bibitem{Prokhorov-etal-2002}
 Prokhorov, M.E., Popov, S.B., and Khoperskov, A.V. 2002, \textit{Astron. Astrophys.}, \textbf{381} 1000--1006

\bibitem{Ruffert-1999}
 Ruffert, M. 1999, \emph{Astron. Astrophys.}, \textbf{346}, 861--877

\bibitem{Ruzmaikin-etal-1998}
 Ruzmaikin, A.A., Sokolov, D.D., and Shukurov, A.M. 1998, \textit{Nature}, \textbf{336}, 341--347

\bibitem{Shakura-1973}
 Shakura, N.I. 1973, \textit{Soviet Astronomy}, \textbf{16}, 756--762

\bibitem{Shvartsman-1971}
 Shvartsman, V.F. 1971, \emph{Soviet Astronomy}, \textbf{15}, 377--384

\bibitem{Stumik-etal-2011}
 Strumik, M., Ben-Jaffel, L., Ratkiewicz, R., Grygorczuk, J. 2011, \emph{Astrophys. J.}, \textbf{741}, 6

\bibitem{Walder-etal-2011}
Walder, R., Folini, D., Meynet, G. 2011, \emph{Space Sci. Rev.}, \textbf{125}, in press

\bibitem{Zeldovich-Shakura-1969}
 Zel'dovich, Ya.B., Shakura, N.I. 1969, \emph{Soviet Astronomy}, \textbf{13}, 175--183
\end{thebibliography}
\end{document}